# Characterizing GPROF Regional Bias Using Radar-Derived Hydrometeor Information

Eric M. Goldenstern, Christian D. Kummerow

*Abstract*—Current satellite precipitation retrievals like GPROF assume that brightness temperature is sufficient to constrain rainfall. This information, however, often represents multiple rain states, resulting in rainfall estimate uncertainties. These uncertainties, while dominated by random variability, can also exhibit substantial regional biases, complicating the use of traditional ground validation techniques which seek to understand these uncertainties. This study aims to characterize the physical contributors to these biases for use in uncertainty quantification. To do this, coincident GPROF Version 7, GMI, and GPM Combined observations were examined over three tropical land regions, the Amazon, Congo, and Southeast Asia, which are known to exhibit distinct biases relative to one another when comparing GPROF with GPM Combined. Rain intensity and ice-rain ratio were identified as the primary sources of GPROF regional biases. By incorporating the information from these sources, the self-similarity between these three regions was brought from within 13 percent to within 7 percent, reducing the interregional bias by half. Including a third constraint based on the polarization-corrected 37-GHz brightness temperature further improved this self-similarity to within 4 percent by accounting for second-order hydrometeor profile differences which were underutilized by GPROF. Comparing the effects of these three constraints between GPROF Version 7 with the 1D version of GPROF-NN showed similar improvements, indicating the utility of this uncertainty quantification and adjustment method across precipitation products. With these constraints, regional GPROF biases can be made more consistent, improving the fidelity of the precipitation climate data records and operational precipitation products which utilize this information.

## I. INTRODUCTION

PRECIPITATION is an essential aspect of the Earth system owing to its role as a controlling factor in the global water and energy cycles. Much of this information is obtained from spaceborne platforms, which circumvent many of the issues confronted by ground-based measurements such as terrain blockage and inhomogeneous spacing [1], [2]. Precipitation is also a highly variable process, both occurring at very fine spatiotemporal scales [3] and being sensitive to many difficult to quantify environmental characteristics and interactions. This variability results in uncertainties in spaceborne precipitation measurements, which then have downstream impacts on climate records of precipitation and in operational product fidelity. While uncertainties exist on the global scale, regional biases have also been noted, further complicating their interpretation from ground validation sites. Given the necessity to provide the most accurate assessments of precipitation possible, efforts to quantify random and systematic uncertainty remain at the forefront of precipitation research.

To date, most uncertainty quantification methods come in the form of either ground validation studies or product intercomparisons. Ground validation studies, which utilize direct comparisons between a high-quality "ground truth" dataset and the precipitation product in question, have served as the backbone for product assessment in most precipitation measurement endeavors. Some examples of this work are detailed in [4]-[6]. These studies, however, are typically performed at local to regional scales, meaning they are not guaranteed to represent the uncertainties in other areas [7]-[9]. This is most impactful for developing areas, where these products are most urgently needed due to the lack of ground truth data [10]. Precipitation product intercomparisons, in contrast, determine the systematic errors of a given product relative to the dispersion of an ensemble of related products [11], [12]. While more readily applied to global precipitation records, such intercomparisons represent errors in terms of statistics as opposed to physics, which can result in significant underestimates or overestimates of uncertainty when the product ensemble makes similar but incorrect assumptions across all members.

More recently, efforts to relate uncertainty to specific kinematic and thermodynamic processes have garnered increased attention. One example focused on climate scale tropical ocean precipitation is from Leitmann-Niimi [13]. This study found that water budget closure, defined as a balance of water vapor divergence, evaporation and precipitation, is mostly affected by precipitation errors that can be related to convective organization and various teleconnection patterns. Another study by Petkovic and Kummerow [14] looked at regional biases in the Goddard Profiling Algorithm (GPROF) over two tropical land regions. Here, it was found that some large-scale thermodynamic and kinematic parameters like Convective Available Potential Energy (CAPE) and surface dewpoint depression showed some utility in characterizing

This work was supported by NASA grants 80NSSC19K0680 and 80NSSC22K0604. *(Corresponding Author: Eric Goldenstern)*

E. M. Goldenstern is with the Department of Atmospheric Science at Colorado State University, Fort Collins, CO 80523 (email: eric.goldenstern@colostate.edu).

C. D. Kummerow is with the Department of Atmospheric Science at Colorado State University, Fort Collins, CO 80523 (email: christian.kummerow@colostate.edu).

Color versions of one or more of the figures in this article are available online at http://ieeexplore.ieee.org

certain precipitation processes that may be related to retrieval biases. Other studies have also indicated that the large-scale environment is capable of influencing precipitation errors through modification of various related parameters and processes [15]-[17].

While the above uncertainty quantification methods have been put forward to describe precipitation uncertainties, none have attempted to determine the parameters which directly affect the observed brightness temperatures (TBs) used by satellite precipitation algorithms to diagnose precipitation, and therefore its potential biases. Since these TBs carry information related to vertical hydrometeor structure, which can be gathered from radar observations, an opportunity is provided to determine which hydrometeor characteristics are most influential to regional biases. Therefore, it is hypothesized that by identifying the primary hydrometeor information which control these TBs, direct diagnosis of precipitation uncertainties can be developed and their regional variations can be explained and accounted for. These information would also provide a pathway by which various kinematic and thermodynamic processes can be related to precipitation estimates, eventually allowing for the diagnosis of bias at various scales. While this study's analyses are focused on Version 7 of GPROF (hereafter GPROF [18]) to determine these hydrometeor characteristics, it is believed that in other algorithms which use similar TB information, these structures will act similarly and thus produce roughly consistent behaviors.

This article is organized as follows. Section 2 describes the data and GPROF retrieval procedure used in this study. Section 3 describes the identification of the primary radar-derived hydrometeor characteristics which can constrain retrieval bias and their effects on this constraint. Section 3 also describes efforts to further improve bias constraint by investigating potential second-order effects. Section 4 describes efforts to compare the effects of these constraints' applicability to other retrieval types. Section 5 provides a summary of the results discussed in the body of the article.

## II. DATA AND METHODS

*A. GPROF Precipitation Retrieval and A-priori Database*

This study employs GPROF as the target precipitation product, which utilizes passive microwave (PMW) information from the Global Precipitation Measurement (GPM) Mission satellite constellation to generate globally available precipitation estimates. GPROF was chosen because it is developed in-house and therefore its architecture and instabilities are more readily accounted for. It also serves as the base algorithm for the Integrated Multi-satellitE Retrievals for GPM (IMERG; [19]) workflow, which has a much larger user base and requires better uncertainty information. To make its precipitation estimates, GPROF relies on Bayes' Theorem, utilizing an a-priori database of TBs and ancillary data such as surface temperature to compare against observed conditions. These comparison are used to develop weights for the different entries, which are then recombined in a weighted average to create a posterior precipitation rate estimate. The GPROF a-priori database is comprised of information gathered from October 2018 to September 2019 from the GPM Combined Radar-Radiometer Algorithm (CMB; [20]), which uses an optimal estimation approach to blend information from the GPM Microwave Imager (GMI) and Dual Precipitation Radar (DPR) and generate physically consistent precipitation estimates. This database will also comprise the data utilized in the subsequent analyses, and both the GPROF and CMB data are available from the NASA PMM group (https://pps.gsfc.nasa.gov).

This study aims to understand the main hydrometeor characteristics which can explain and constrain GPROF uncertainty. Toward this end, the GPROF algorithm was run in an experimental mode where it was tasked to predict its own database entries. The advantage of this mode is that the database provides an objective truth in the database rain rates by which to compare the GPROF estimate to. This experimental mode also avoids ground validation data that often has its own set of issues that must be dealt with when comparing different locations. Validating against an independent database maintains the focus on what GPROF itself and is thus more appropriate for determining intrinsic algorithm errors. Care was also taken to avoid including the database entry being estimated from inclusion in the a-priori database by identifying the entry with the highest Bayesian weight and excluding it from the estimation phase.

*B. GPM Combined Hydrometeor Information*

To determine the effects of various sources of hydrometeor information, the rain and ice water profiles from CMB were used. These profiles provided water content information up to 18km above ground level (AGL) and were taken from the CMB data which had been averaged into the GPROF footprint. This information is directly linked to the GPROF precipitation estimates, as these profiles are used as input in a forward radiative transfer model to generate the TBs in the a-priori database. Though cloud liquid water profiles were also available, they were not treated as an observable. Before these rain and ice water profiles were used in the subsequent analyses, some modifications were performed to improve the physical consistency between the radar and TB observations. Due to differences in the scan geometry between DPR (nadir-pointing) and GMI (52-degree incidence), the water content profiles were matched to the same footprint location in the scan line immediately preceding a given GPROF footprint. This was done to better align the profiles with the TBs used by GPROF in its estimates. The profiles were also adjusted in the 2km deep layer centered at the freezing level using a linear decay for rain and a linear growth for ice to reduce the effects of bright banding.

From these profiles, characteristics such as water path and layer water content were developed. The path variables ($Path$) were calculated as a summation between the product of the individual level water contents ($WC_i$) and the distance between that level and the previous one ($\Delta z_i$) through the entire profile extent (j = 28 levels), as shown in (1). Layer water content ($WC_{LYR}$) was defined as the sum of the individual level water contents within a given depth ($L$), as shown in (2).

$$Path = \sum_{i=1}^{j} WC_i \Delta z_i \qquad (1)$$

$$WC_{LYR} = \sum_{i=l}^{L} WC_i \qquad (2)$$

*C. Study Database and Uncertainty Quantification Method*

To focus on the regional variability in GPROF uncertainty, the a-priori database used in the experimental GPROF mode was restricted to three regions of interest (ROIs) within the Tropics (Fig. 1) and to entries associated with GPROF surface classes 3 (maximum vegetation) and 17 (mountain rain). The choice to limit this study to these regions and surface classes was made to minimize the variability introduced by different climatic regimes and surface conditions. The ROIs chosen include the Amazon (AMZN), Congo (CNGO), and Southeast Asia (SEAS). AMZN and CNGO were defined identically to those in Petkovic and Kummerow [14], while SEAS was created as an additional test region within the boundaries of 5-25N and 90-110E. These ROIs were chosen to represent tropical landmasses which experience large amounts of annual rainfall but are known to have different biases in GPROF. For the uncertainty analyses, this subset was further reduced to only those entries where the CMB rain rate was between 0.25 and 32 mm/hr. This reduction was decided based on the range of rain rates which represented at least 95 percent of the accumulated rainfall in the restricted database. While the "extreme" precipitation rates are still important, the purpose of this study was to identify the typical behavior of GPROF and the associated hydrometeor information, which is better served by mitigating the interference from outliers.

The metric chosen to quantify GPROF biases in this study was the bias ratio. As shown by (3), *Bias Ratio* was created as the ratio of the accumulated precipitation from the retrieval ($\sum RR_{GPROF}$) to the accumulated precipitation from the truth dataset ($\sum RR_{CMB}$).

$$Bias\ Ratio = \frac{\sum RR_{GPROF}}{\sum RR_{CMB}} \qquad (3)$$

Though CMB is itself an estimate, it is radar derived and therefore closer to an objective truth than GPROF, particularly over land where PMW information content is lower. A bias ratio of 1 represents an unbiased measurement, while ratios less than 1 are underestimates by GPROF and ratios greater than 1 are overestimates. Another benefit of using bias ratio is that it can be inverted to remove the associated bias from GPROF through a simple multiplication of the two. Such an adjustment allows for both an objective quantification of bias and a method by which to determine the contribution of a specific phenomenon to the overall systematic bias. Initially, this adjustment was used to scale the reduced GPROF database to ensure it was unbiased overall. This choice was made to ensure that focus was given to the regional discrepancies as opposed to artificial biases introduced by restricting the database.

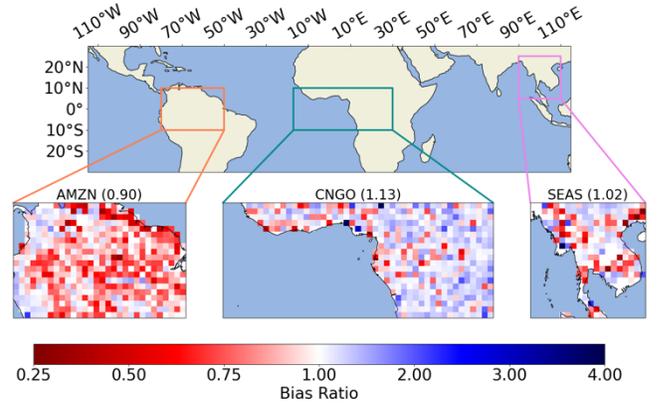

**Fig. 1.** The three regions of interest (ROIs) utilized in this study. The values in parentheses are the ROI mean bias ratios, while the colored grids are the 1-degree ROI bias ratios.

### III. RESULTS

*A. Initial GPROF Regional Bias Analysis*

Since GPROF is unbiased globally, the biases by ROI were considered. To investigate these biases, two sets of bias ratios were developed for each ROI: an overall bias using the full set of ROI data and 1-degree gridded bias ratios using the ROI data contained by each grid space. In this way, an investigation of both the general impact of biases on GPROF and their spatial variability within the ROIs was possible. Fig. 1 displays both perspectives. First considering the overall biases, AMZN and CNGO exhibited obvious systematic differences, being 0.9 (10 percent underestimation) and 1.13 (13 percent overestimation) respectively. These biases were consistent with those found in Petkovic and Kummerow [14]. SEAS exhibited the lowest overall bias of the ROIs at 1.02, or 2 percent overestimation, which would indicate that this ROI is relatively well-represented by GPROF. From these bias ratios, the three ROIs considered agree to within 13 percent of each other, highlighting the large regional discrepancy in the regional uncertainties.

Including the spatial distribution of these biases adds additional context to these overall biases. Looking first at AMZN, the systematic underestimation resulted from broad underestimation across the entire region. This underestimation was most concentrated along the northeastern coast of the landmass and within the rainforested portions of the ROI. These areas typically experience precipitating systems that are initiated by coastal squall lines, which due to the maritime nature of the associated airmass are often shallow and dominated by warm-phase microphysics [21]. This was contrasted by an overestimation tendency located primarily over the Andes Mountains, which often serve as a focal point for generating deep convection.

The bias spatial characteristics in CNGO contrasted with AMZN through the relatively uniform overestimation throughout the ROI. This region is known for having some of the most vigorous deep convection in the world, driven largely by the African easterly jet and some enhancement from the Ethiopian Highlands. While CNGO was generally overestimated, some areas of underestimation appear along the west-central and northwestern portions of the ROI landmass. These locations are roughly located with the Intertropical Convergence Zone (ITCZ) and often experience westerly Atlantic flow, resulting in more maritime precipitation characteristics [22].

SEAS exhibited much larger spatial variability in its biases than both AMZN and CNGO, likely explaining its relatively low overall bias. This region is topologically and climatologically complex, with seasonal impacts from the Asian Monsoon and Madden-Julian Oscillation (MJO) interacting with the Annamite and Luang Prabang mountain ranges [23], often resulting in enhanced deep convection along these mountains and stronger maritime influences over the lowland and coastal areas of the ROI. This again seems to line up with the spatial distributions of bias seen in AMZN and CNGO, where the areas of precipitation more closely related to deep convection are overestimated and those with stronger oceanic influences are characteristically underestimated.

*B. Main Constraint Identification and Partitioning*

To first determine the primary sources of the above GPROF regional biases, its known instabilities and preferred information sources were considered. As a Bayesian algorithm, GPROF creates an average state by combining multiple database entries with similar TBs and develops its precipitation estimate as a weighted average of these states. This rather simplistic methodology results in all estimates made by GPROF being driven towards the mean of the a-priori database. Such a tendency results in the systematic overestimation of lighter rain rates and underestimation of higher rain rates, as shown in Fig. 2. This makes rain intensity the most prominent source of bias in GPROF. This also generally explains why GPROF is unbiased overall, as the bulk of the precipitation accumulation histogram in Fig. 2 resides in rain rates which were relatively unbiased when taken in aggregate. This would imply that rain rate drives GPROF regional biases due to that region's preference for certain rain intensity ranges. To avoid impacts from ice water contamination and near-surface effects, a reference rain rate (*refRR*) was used to quantify the impacts of this bias source. This quantity was calculated using the layer rain water content from a 1-km deep layer centered 1.5km below the freezing level (*refRWC*). This rain water content was converted into *refRR* using (4), which can be derived using the Marshall-Palmer distribution [24] and raindrop fall velocities. This conversion only serves to interpret biases in terms of rain rate and otherwise does not impact any results.

$$refRR = 1.7 * (refRWC^{1.19}) \quad (4)$$

The PMW information content available to GPROF provides another clue for bias constraint. Over land, the high and variable emissivity of the land surface is nearly indistinguishable from the liquid emission of cloud and rain water, leaving ice scattering as the largest source of information content. This scattering is observed in high frequency PMW channels like the 89- and 166-GHz channels in GMI, where TBs are reduced as ice water path, and therefore scattering, increases [25], [26]. This information is often associated with increased precipitation, which results in GPROF prescribing higher precipitation values to these conditions. This increased ice amount, however, is more closely related to cloud microphysical processes and need not result in increased precipitation. As a result, GPROF footprints with greater ice amounts are often overestimated and those with little to no ice underestimated. As such, ice amount is a potential bias source, assuming this characteristic varies systematically between regions. This effect also couples with rain intensity, since a given rain rate can be associated with multiple ice amounts. As such, the amount of ice water with respect to the amount of rain water, here defined as ice-rain ratio (*IRR*), is another likely driver of GPROF biases. For our purposes, *IRR* was defined in (5) as a ratio between the ice water path from 1.5km above the freezing level upward (*IWP*) and *refRWC*.

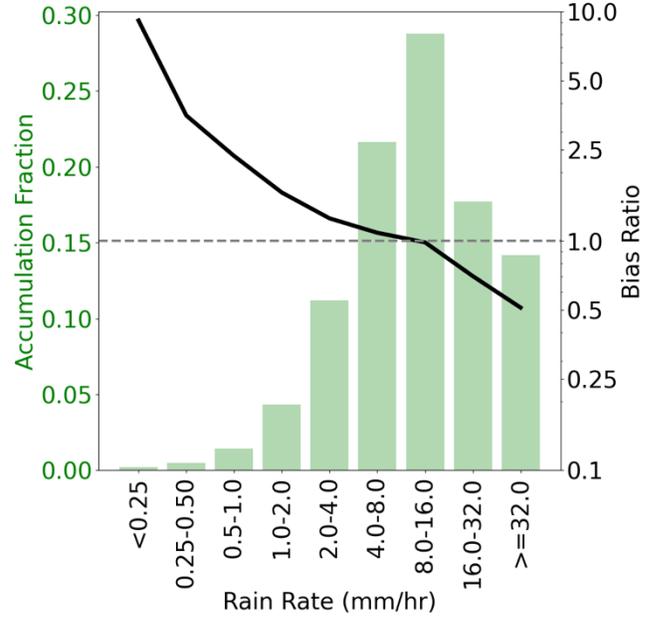

**Fig. 2.** The rain rate accumulation histogram (green bars) and associated bias ratios (black line) for the restricted database.

$$IRR = \frac{IWP}{refRWC} \quad (5)$$

Fig. 3 shows how bias ratio varies by IRR. As IRR was increased, the bias ratios shifted from underestimation to overestimation. This broadly follows the results from Braga and Vila [27], which found that rain rates in precipitation retrievals often increase alongside ice amount. The underestimation at lower IRR values also highlights well-known issues in GPROF regarding warm-phase precipitation, as described by Ryu et al. [28], among others. This relationship between IRR and bias ratio also showed sensitivity to rain intensity, with the highest rain rates remaining underestimated at relatively large IRR values while low rain rates were overestimating at relatively low IRR values. This indicates that IRR should be used alongside rain intensity to characterize its effects on bias.

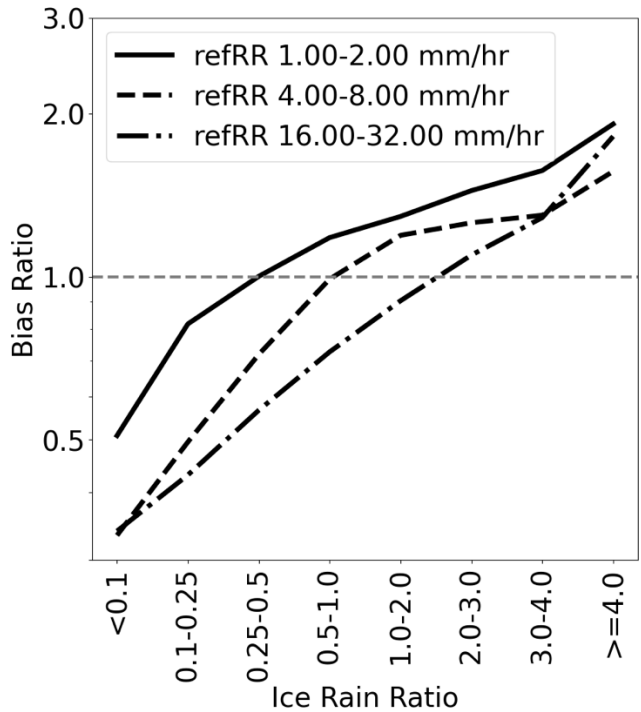

**Fig. 3.** The bias ratios from the reduced database as a function of ice rain ratio. Each line represents the data attributed to varying rain intensities.

It is also worth noting that because refRR is being used to compute bias ratio instead of surface rain rate, the vertical distribution of rain water may also impact biases. This rain water distribution, hereafter referred to as RRslope, would directly relate to biases due to its ability to describe how well refRR represents surface processes. In cases where RRslope indicated preferential water collection at the reference level (high positive slope), one would expect overestimation to occur and vice versa where the distribution preferentially collected water near the surface (high negative slope). For this study, RRslope was determined as the percent change between the reference ($refRWC$) and near-surface ($sfcRWC$) layer water contents, as shown in (6). The lowest three levels of the rain water profile were used to calculate sfcRWC via (2). While this characteristic was recognized as another potential bias source, it can only provide information if the different regions exhibit separable preferences for these slopes. This will later be shown to not be the case for these three ROIs, since they all represent similar conditions as tropical landmasses.

$$RRslope = \frac{refRWC - sfcRWC}{sfcRWC} \quad (6)$$

To look at how these parameters can be used to interpret GPROF biases, the precipitation data were grouped based on refRR and IRR. Eight bins of IRR and nine bins of refRR, as shown in Fig. 4, were used to create a total of 72 groups. These groups were also later divided using three bins of RRslope, determined by that parameter's terciles. The refRR and IRR bins were determined by hand for refRR (0.25 and 32 mm/hr) and IRR (0.1 and 4). These ranges represent 95 percent of their respective distributions, separating the "extreme" cases from the main behavior in the database. The bin intervals were first created by considering the smallest possible number of partitions which were believed to capture roughly similar environmental conditions. If a given partition experienced a notable decrease in the standard deviation of the bias ratios it represented when split in half, these new bins were retained. Care was also taken to ensure the bins were sufficiently populated, defined here as containing at least 1000 entries, to ensure they remained robust.

Fig. 4 also displays the mean bias ratios associated with each bin. From this figure, several conclusions can be drawn regarding GPROF's bias tendencies. Both the behaviors discussed regarding bias ratio as a function of refRR and IRR are clearly visible. This explains the strong overestimation found at high IRR and low refRR values and vice versa at low IRR and high refRR values. As IRR increases, the overestimation bias is extended into the moderate refRR bins. This speaks to GPROF's reliance on ice scattering information, with higher IRRs inflating the rain rates beyond their expected values. Once refRR is sufficiently large, however, the underestimation tendency dominates even at high IRR values. This speaks to the rain intensity bias in GPROF, where preferential overestimation was occurring at low rain rates and underestimation at high rain rates.

Using the bias ratios shown in Fig. 4, their inverses were determined and used to adjust the GPROF precipitation by group. In this way, the biases contributed by both refRR and IRR were removed from the data, leaving a residual bias to compare against the original bias. If these hydrometeor characteristics can fully explain GPROF biases, then the residual will indicate an unbiased result. The effects of this adjustment on the ROI biases are shown in Fig. 5. First looking at the mean bias ratios by ROI, substantial improvements were noted. As opposed to the ROIs agreeing to within 13 percent (0.90-1.13) in Fig. 1, they now agree to within 7 percent (0.95-1.07), representing an increase in regional similarity of nearly 50 percent. This improvement shows that by describing the underlying hydrometeor structure, additional information can be provided which aids in the interpretation of GPROF biases. The overall biases were also drawn closer to 1, indicating that accounting for the different conditions represented by refRR and IRR improved the correspondence between GPROF and CMB, showing the viability of using these bias indicators to potentially improve GPROF biases. Adding RRslope to this refRR-IRR based adjustment yielded nearly the same results as refRR-IRR alone, suggesting that this parameter does not contribute meaningful information for reducing the original biases. This is believed to be related to these ROIs having very similar low-level moisture environments, which may show random variations in RRslope but not the systematic differences required to describe their regional biases.

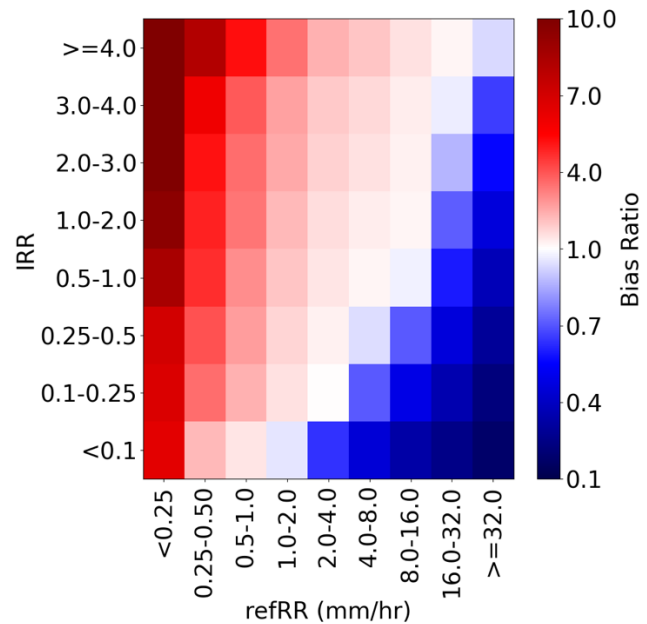

**Fig. 4.** The bias ratios for each of the 72 reference rain rate (refRR) – ice rain ratio (IRR) groups. These ratios are determined as the mean ratio for each group.

Despite these improvements in the overall self-similarity of the regions, the systematic differences in bias mirrored those before adjustment, as AMZN was still categorically underestimated by 5 percent and CNGO and SEAS categorically overestimated by 7 and 1 percent respectively. These systematic difference were also observed in the spatial bias distributions shown in Fig. 5, where many of the same structures in Fig. 1 remained visible. In AMZN, categorical underestimations over the rainforest and along it northeastern coastline alongside overestimation associated with the Andes were still present. CNGO remained systematically overestimated, though more prominent underestimation tendencies were noted in areas associated with ITCZ convection and westerly tropical Atlantic flow. SEAS appeared to have stronger underestimation in the lowland and coastal areas while maintaining, if slightly reducing, the overestimation in the northern mountainous areas. From these results, it was therefore noted that while controlling the regional bias using refRR and IRR resulted in much-improved overall agreement, additional constraints still exist which can describe the remaining systematic biases.

*C. Second Order Sources*

Using IRR and refRR, roughly 50 percent of the GPROF bias can be explained. Given that these two parameters represent the most likely first order constraints on bias and RRslope did not appear to contribute additional information, the existence of second order bias sources was investigated. Since the primary information used by GPROF resides in the high frequency channel TBs, the polarization-corrected 37-GHz brightness temperature (PCT37) was examined as a potential bias source. PCT37 is high affected by surface emissivity but is capable of contributing information regarding both liquid and ice hydrometeors under certain conditions such as large overall water contents and enhanced concentrations of sizable ice particles [29], [30]. For this study, PCT37 was calculated following Cecil and Kronis [31].

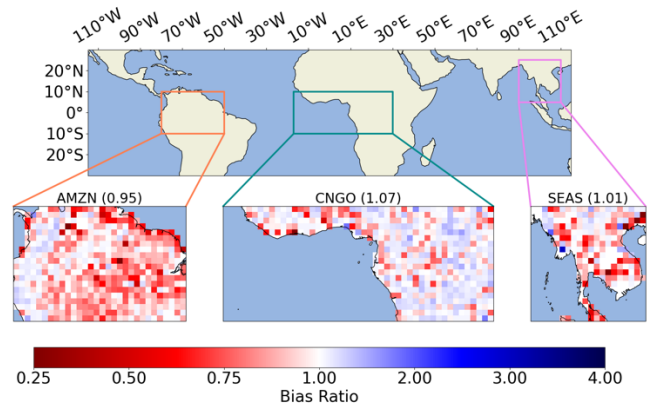

**Fig. 5.** The same as Fig. 1, but after refRR-IRR adjustment.

Fig. 6 shows the bias ratios for select refRR-IRR groups as a function of PCT37. From this figure, colder PCT37 TBs were generally associated with overestimation and vice versa for warmer PCT37 TBs, with the most notable changes occurring at the warmer end of the range. This behavior fits with that expected from the high frequency channels, where colder TBs are associated with greater ice water paths which often represent overestimations. This behavior, however, appeared strongly affected by the refRR-IRR group being considered. For example, using the groups in Fig. 6, a PCT37 TB of 270 Kelvin could have a bias ratio between 1.5 and 4 depending on the group being considered. This difference by group, along with the fact that bias ratio did not show particularly strong sensitivity to PCT37 on its own, suggests that this constraint is most useful when used complementary to the refRR-IRR groups.

To incorporate the information provided by PCT37 into the previous analyses, this parameter was partitioned into six bins using the endpoints of 265 and 285 Kelvin and bin widths of 5 Kelvin. These bins were developed using similar decisions as those for refRR and IRR. The refRR-IRR groups were then broken down by the PCT37 bins, creating a total of 432 refRR-IRR-PCT37 groups from which to evaluate GPROF biases. The bias ratios were then recomputed for this new set of groups. Fig. 7 shows the results of recomputing biases in the refRR-IRR space for two of the PCT37 groups. From this figure, many obvious differences exist compared to the biases shown in Fig. 4. First considering the refRR-IRR bias ratios in the 265-270 Kelvin PCT37 bin, it was apparent that the overestimation tendency extended through a much greater portion of the data, with only those cases with very heavy rain rates being underestimated. There was also a stronger preference towards the rain rate component of the bias, as the gradient change across refRR was overall greater than that for IRR. Looking at the 280-285 Kelvin PCT37 bin, a roughly similar picture as that in Fig. 4 existed, but with some noteworthy differences. Like in the 265-270 Kelvin bin, the refRR bias component was more pronounced, with an even stronger bias gradient across the refRR bins. The sensitivity of bias ratio to IRR also appeared to decrease, both in regard the bias gradient along the IRR bins and its ability to extend GPROF's overestimation tendency into higher rain rates. Compared to Fig. 4, high values of IRR no longer resulted in overestimation at more moderate rain rates, further reinforcing the dominance of the rain rate bias under these conditions. Given the differences noted both between these PCT37 bins and compared with the bias ratios shown in Fig. 4, including PCT37 in the bias ratio determination provided additional information to the refRR-IRR groups based on the PCT37 bin being considered.

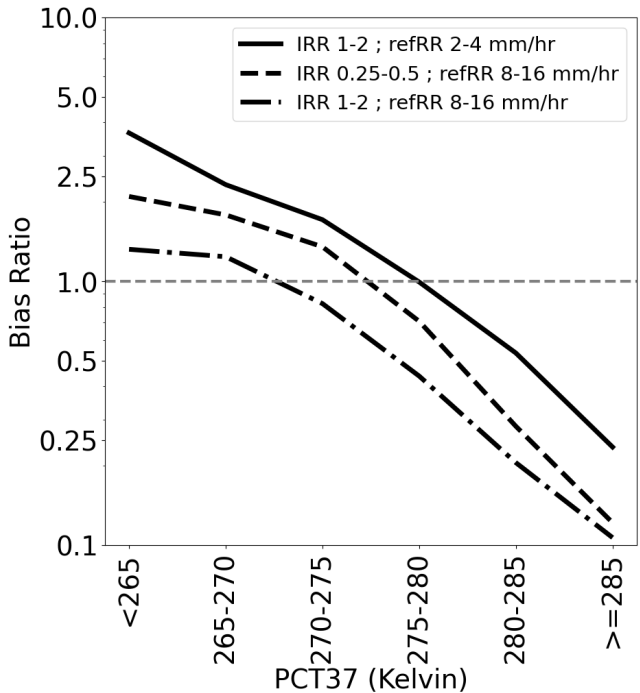

**Fig. 6.** GPROF bias ratios as a function of polarization-corrected 37-GHz brightness temperature (PCT37) for three representative refRR-IRR groups.

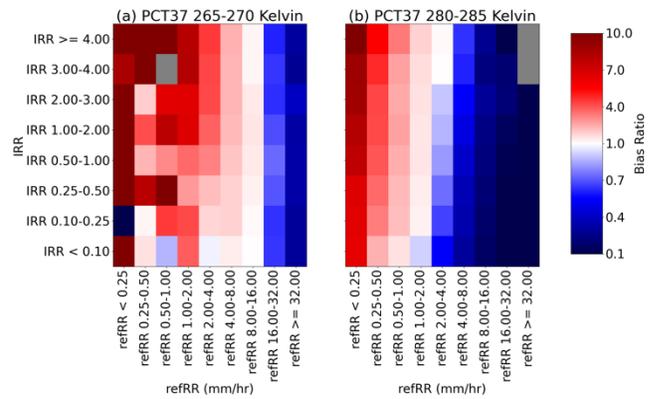

**Fig. 7.** The same as Fig. 4, but incorporating the (a) 265-270 Kelvin and (b) 280-285 Kelvin polarization-corrected 37-GHz (PCT37) bins. Gray boxes indicate groups with no entries.

Like in the previous analysis, the inverses of these new bias ratios were used to adjust the associated GPROF precipitation estimates and the residual bias was recalculated. These results are shown in Fig. 8. By incorporating PCT37 into the refRR-IRR groups, the agreement between the ROIs improved to within 4 percent (0.96-1.03), showing a nearly 40 percent improvement from the refRR-IRR group adjustment. AMZN and CNGO were brought even closer to an overall bias of 1, further indicating the utility of PCT37 information to further improve on GPROF biases. SEAS, however, notably shifted from an overestimation to an underestimation tendency. Despite this shift, the region remained comparable with AMZN and CNGO, so this behavior was attributed to the higher variability in precipitation within this ROI.

Looking at the spatial distribution of bias ratio by ROI in Fig. 8, it was apparent that this new grouping not only further reduced the bias ratio magnitudes compared to the refRR-IRR adjustment but better accounted for the systematic differences in spatial structure noted in both Figs. 1 and 5. Comparing AMZN and CNGO, for example, the ROIs showed much reduced preferences for underestimation and overestimation, respectively, instead showing a more heterogeneous mix of the two. Portions of the rainforest within AMZN now display mostly neutral biases and even some areas of overestimation, while larger portions of CNGO now show underestimation tendencies where they previously were overestimated. This, along with the increased spatial heterogeneity of the biases, gave an appearance that the residual bias was more representative of random error than systematic error. One interesting feature in AMZN was that the Andes had become a source of underestimation. This behavior over mountainous regions also appeared in the mountainous areas in SEAS, suggesting that orography has additional behaviors that are not correctly accounted for in this method.

*D. Physical Links between PCT37 and Bias*

While using PCT37 alongside refRR and IRR to explain GPROF regional bias greatly improved on the refRR-IRR adjustment alone, the exact mechanism by which this additional constraint affected the bias was unclear since the individual 37-GHz channels are already included in GPROF. Two potential explanations for PCT37's usefulness were considered: PCT37 provided information which the individual 37-GHz channels did not capture or GPROF underutilized the information provided by the 37-GHz channels. Since PCT37 is not a direct input to GPROF, but rather attempts to provide a surface-corrected proxy for the 37-GHz channels, it is possible that this parameters adds surface information that is lost in the noise of the individual channels. Additionally, it is possible that the individual 37-GHz channels do discern the same information as PCT37, but because GPROF is tuned to focus on the higher frequency channels it tends to disregard this information.

To first see if PCT37 was contributing information which was not apparent in the individual 37-GHz channels, mean rain and ice water profiles were developed for each of the PCT37 bins. These profiles were also developed by refRR-IRR data group since PCT37 showed its most substantial contributions when used in tandem with these two parameters. The individual 37-GHz GMI channels were similarly partitioned to create their respective mean profiles. These profiles were then qualitatively compared to see how the different PCT37 and 37-GHz bins separated different vertical hydrometeor structures. Fig. 9 shows an example of these profiles for the IRR 1-2 and refRR 8-16 mm/hr group, which represented about 10 percent of the total database rain accumulation. These profiles were also developed for each ROI and the other refRR-IRR groups and were found to behave like those in Fig. 9. From this figure, it was apparent that the individual 37-GHz channels showed very similar mean hydrometeor profile structures as PCT37 across all bins, indicating that PCT37 was not providing new information compared to the individual 37-GHz channels. This means that GPROF does have this information, and therefore must not be using it to its fullest potential.

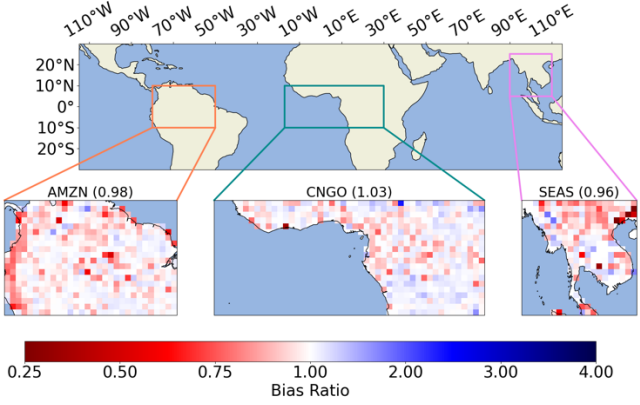

**Fig. 8.** The same as Fig. 1, but after the refRR-IRR-PCT37 adjustment.

Since the 37-GHz information appears to be underutilized by GPROF, further consideration was given as to why this is the case. While the full picture of this underutilization remains a topic of discovery, some clues have been found in the previous analyses which may explain this behavior. One clue is that PCT37 shares information with the higher frequency channels. Individually, the 37-GHz channels can react to both changes in ice scattering and liquid emission. Over land, the ice scattering component is dominant, and like in the high frequency channels, results in decreased TBs as ice water path increases. This means that the 37-GHz channels react to changes in hydrometeor characteristics much in the same way as the high frequency channels. Fig. 9 hints at this, with increases in ice water path with decreasing PCT37, much like what would happen for the 89- or 166-GHz channels. The high frequency channels, however, are more sensitive to ice scattering than the 37-GHz, making them the stronger information source. This means the major contribution of 37-GHz is no longer unique, resulting in the information it provides no longer being separable by GPROF.

Once the refRR-IRR groups are introduced, however, PCT37 showed an ability to separate data with different bias characteristics. This seemed to be tied to PCT37's ability to identify subtle differences in the rain and ice water profiles, shown in Fig. 9, to better organize the entries within each refRR-IRR group. Some of the most prominent differences are noted in the distribution of low level rain water, much like that described by RRslope. At low PCT37 TBs, the rain water profile prefers a relatively constant water content with height, while at high TBs this shifts to preferred rain water collection near the freezing level. There are also notable differences between PCT37 bins within the mixed-phase region, here defined as the layer between 5 and 7km AGL. Here, the mean profiles appeared to favor decreased ice water and increased rain water when PCT37 was warm and vice versa when PCT37 was cold, though ice water remained the dominant signal. Even above the mixed-phase region, the shape of the ice water profiles differed considerably, with the warmest PCT37 bin ice water being most heavily concentrated in the mixed-phase region while the coldest PCT37 bin displayed a stronger ice profile in the upper atmosphere. As such, the function of PCT37 appeared to be organizing the entries in each refRR-IRR group by these subtle differences. Since the hydrometeor profiles are more directly linked to precipitation, this allowed for better description of the precipitation estimates, and therefore their biases.

IV. COMPARISON WITH GPROF-NN

Having identified refRR, IRR, and PCT37 as the most likely constraints on GPROF regional bias, some questions remained regarding the applicability of these sources to other precipitation retrievals. While refRR and IRR are likely to retain their degree of importance given the well-established theories behind them, PCT37's utility may be a result of the GPROF algorithm architecture and therefore may not be applicable to other algorithm types. To test this, we considered an additional retrieval: Version 8 of GPROF, also known as GPROF-NN, which utilizes a quantile regression neural network (QRNN) approach that has been shown to improve on both GPROF's retrieval accuracy and computational efficiency. GPROF-NN exists in two architecture types: a point-based 1-D type and a 3-D type which incorporates spatial context via the use of convolutional layers. Further information on both GPROF-NN algorithms is presented in Pfreundschuh et al. [32].

To investigate the applicability of these bias constraints for GPROF-NN, the previous bias analyses were reproduced using the 1-D GPROF-NN as the target retrieval. The 1D version was chosen because it uses single pixel information, making it the more comparable architecture to GPROF. The GPROF-NN dataset used for these analyses was also restricted and scaled using similar methods as for GPROF. It is important to note here that while the above steps were taken to match the GPROF-NN data to that for GPROF as closely as possible, some differences between the two retrieval could not be removed. Specifically, the database used for GPROF-NN is not restricted to the three ROIs as it was for GPROF. This crucial difference, among others like GPROF-NN's inclusion of an orographic adjustment, means that the comparison being carried out is not exact. Rather, this exercise was designed to investigate if the bias sources identified in GPROF can produce similar behaviors when used with another algorithm.

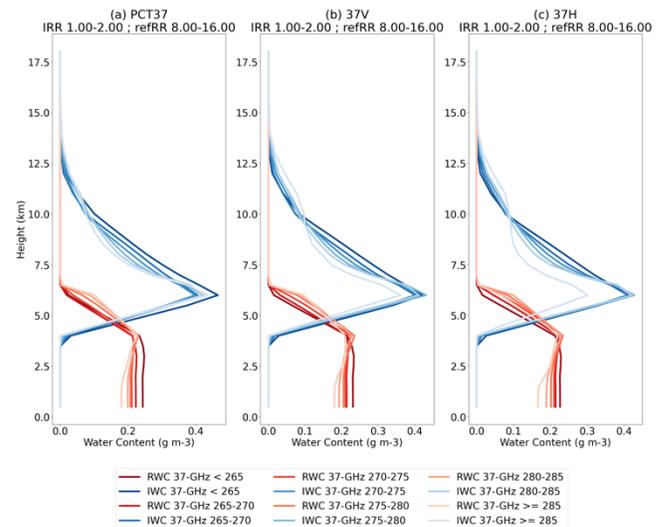

**Fig. 9.** Mean rain and ice water profiles by bins of (a) PCT37 and the (b) vertically polarized 37-GHz and (c) horizontally polarized 37-GHz. These profiles were developed for the IRR 1-2 ; refRR 8-16 mm/hr group.

GPROF-NN can be shown to have similar, though muted, biases as GPROF with respect to both refRR and IRR, so the focus of this comparison is whether adding PCT37 to the refRR-IRR group information provides similar improvements in GPROF-NN as it did with GPROF. As previously discussed, changing between these groups in the GPROF data prompted an improvement in regional self-consistency from within 7 percent to with 4 percent. In the case of GPROF-NN, this same change improved the consistency between the ROIs, with some caveats. When considering only AMZN and CNGO, the GPROF-NN bias consistency improved from within 3 percent to within 1 percent, representing remarkably similar amounts of improvement compared with GPROF. SEAS, however, was dominated by the mountain class which was affected heavily by the previously described orographic adjustment. This resulted in a strong overestimation bias remaining present here even after data scaling. Reconfiguring the scaling method to adjust the two GPROF surface classes separately resulted in very similar results as for GPROF in this ROI, shifting its bias ratio from 1.03 to 0.95. While using a different scaling method for GPROF-NN to compensate the orographic adjustment makes interpretation of these results difficult, the three bias constraints still appeared to improve the regional biases and bias consistency in GPROF-NN, signaling their potential utility across algorithms.

## V. DISCUSSION

Uncertainties in GPROF precipitation originate from multiple sources. Some of these sources are algorithmic, being related to the retrieval technique itself, but the bulk of the uncertainty is related to GPROF's inability to separate TBs which occur under different rainfall conditions. This non-uniqueness inevitably results in errors but can have a more physical connection to large-scale atmospheric properties through the vertical hydrometeor structure associated with the TBs. Using bias ratio as an uncertainty quantification metric, several sources of GPROF bias were identified. The first of these sources was refRR, which created biases due to the tendency of GPROF to average multiple precipitation states together when developing its precipitation estimates. This leads to predictable overestimates at low rain rates and underestimates at high rain rates. The second was IRR, which directly related to the ice scattering properties at a given location and comprised the primary signal by which GPROF developed its estimates of terrestrial precipitation. In general, increasing IRR results in increasing overestimation due to GPROF's tendency to inflate rain rates in cases of strong ice scattering. The third was PCT37, which contextualized the information from refRR and IRR by identifying multiple second-order differences in the rain and ice hydrometeor structures but was underutilized by GPROF due to its limited information content. A fourth bias source considered was RRslope, which described the vertical distribution of rain water between the reference level and an analogous surface level.

Having identified these potential bias sources, their exact effects were quantified and accounted for through adjusting the associated GPROF precipitation by the inverse of their corresponding bias ratios. Doing this using the 72 refRR-IRR groups showed an improvement in regional bias consistency between the three ROIs of about 50 percent, with the regions agreeing to within 7 percent as opposed to 13 percent without adjustment. Expanding these groups to include PCT37 further increased this agreement to within 4 percent, marking a 70 percent improvement over the unadjusted GPROF precipitation. When considering just refRR and IRR, a reduction in the magnitude of biases was noted, but the systematic differences between regions remained intact. Including RRslope in the refRR-IRR groups did not yield noticeably different performance, indicating that this parameter does not have a large impact on regional biases in the Tropics. When PCT37 was added, however, the systematic regional differences diminished considerably, indicating that this source greatly improved on the constraints provided by the refRR-IRR groups alone.

Since the bias characteristics of the individual constraint parameters are physically based and location invariant, the ability of these parameters to constrain GPROF regional biases must be related to the varied preferences of each constraint between the three regions. This can be shown using probability distribution functions (PDFs) of the four potential bias sources separated by ROI, as shown in Fig. 10. In refRR, IRR, and PCT37, there were apparent differences in these regional PDFs. More specifically, CNGO showed a greater preference for colder PCT37 TBs compared to AMZN, with their distributions peaking roughly 5 Kelvin apart, while the SEAS distribution fell somewhat in-between the two. Since colder PCT37 TBs were generally associated with an overestimation tendency, this would support the bias tendencies noted for the three regions. CNGO also showed greater preference toward high IRR values (IRR > 1) than the other two regions, which preferred lower ice regimes (IRR < 1). Again, higher IRR represented overestimation, further reinforcing the regional biases described. For refRR, the distribution differences were less distinct, but in general CNGO displayed a preference for enhanced rain rates in comparison with AMZN and SEAS. Lower refRR values were associated with overestimation, so these preferences act opposite of the three regions' biases. Given that the other two constraints support the systematic differences in the ROI biases, refRR likely acts as a mediating factor, somewhat blunting the response. In the case of RRslope, there were very few differences in the regional PDFs outside of the widths of the distributions. As such, this parameter was unable to provide additional context to GPROF bias due to it having similar preferences across these regions. It is possible, however, that including regions with substantially different climatologies, such as desert regions, may show a more distinct separation in the RRslope distributions. As such, the utility of these parameters outside of these three regions should also be considered.

The applicability of these groups to other algorithms was also tested using the 1D GPROF-NN, which swaps the traditional Bayesian approach of GPROF for a neural network approach. By reusing the bin configurations developed for GPROF, similar improvements in the regional bias consistency in the GPROF-NN data were observed, though the interpretation was complicated by differences in the databases and assumptions used to inform the two algorithms. Still, this experiment proved useful for both understanding the innate ability of a retrieval algorithm to distinguish between similar precipitation profiles based on TBs alone and if including information about the underlying hydrometeor structure can aid this distinction.

While identifying these characteristics which better constrain GPROF biases is a crucial step forward in the uncertainty quantification of satellite precipitation retrievals, there are still limits to their application. The most obvious of these limitations is that both refRR and IRR are radar-derived parameters and thus are unobtainable in locations which are not observed by CMB. This is also the case for historical data before spaceborne radars were available and for any current satellites which are not the GPM Core Observatory. As such, these bias constraints cannot be developed for much of the actual GPROF algorithm. The biggest impediment to this is likely IRR, since refRR can be estimated from GPROF and 37-GHz information is generally available. This parameter may have to be determined using large-scale atmospheric properties, regional climatologies, or some combination thereof.

The limited scope of this study also presents potential generalization issues. Since these results were made for three tropical land regions, there is no guarantee that the same results exist in other climatologically different regions such as the Southwest United States or the Siberian tundra. In these cases, the three constraints identified may not adequately describe biases here owing to the presence of other relevant processes. Moreover, additional parameters not used in this study, such as RRslope, may be more impactful in assessing bias across climate types and therefore should be included. To this end, expansion of the work laid out in this study into other, climatologically different locations would provide further information for developing a routine which can be confidently applied everywhere the retrieval is performed.

Regardless of these limitations, this study presents a step towards characterizing and constraining biases in satellite precipitation retrievals through physical arguments. By identifying the physically explainable factors controlling these biases, the context in which they occur is better understood and can therefore be accounted for in a more interpretable manner. This is not only useful for stabilizing precipitation climate data records as retrieval products come and go, but also provides additional methods which forecasters can use to assess the precipitation outlooks they receive. In all, the results of this study show an avenue by which a knowledge base for precipitation uncertainties as they relate to known atmospheric processes can be developed.

## ACKNOWLEDGMENT

GPROF Version 7 experimental code and troubleshooting advice provided by Paula Brown (Colorado State University). GPROF-NN output provided by Simon Pfreundschuh (Colorado State University). Additional advice contributed by Dr. Hernan Moreno and Andres Monsalve Salazar (University of Texas El Paso). The authors also thank the anonymous reviewers for their feedback.

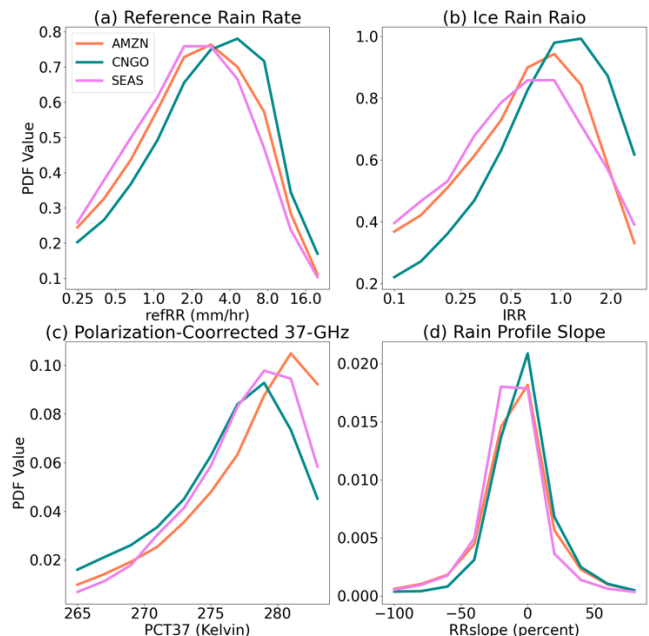

**Fig. 10.** Probability distribution functions for (a) refRR, (b) IRR, (c) PCT37, and (d) RRslope by ROI.